\documentclass[pdftex,pre,twocolumn,groupedaddress,floatfix,showpacs]{revtex4-1} 
\usepackage{graphicx}
\usepackage{url}
\usepackage[utf8]{inputenc}
\usepackage{latexsym,dcolumn,graphicx,amsmath,amssymb}

\usepackage{color}
\usepackage{ifthen}  
\usepackage{changes}

\begin{document}

\title{Dynamics of interacting information waves in networks}
\author{A. Mirshahvalad}
\email{atieh.mirshahvalad@physics.umu.se}
\affiliation{Integrated Science Lab, Department of Physics, Ume{\aa} University, Ume{\aa}, Sweden}

\author{A. V. Esquivel}
\affiliation{Integrated Science Lab, Department of Physics, Ume{\aa} University, Ume{\aa}, Sweden}

\author{L. Lizana}
\affiliation{Integrated Science Lab, Department of Physics, Ume{\aa} University, Ume{\aa}, Sweden}

\author{M. Rosvall}
\affiliation{Integrated Science Lab, Department of Physics, Ume{\aa} University, Ume{\aa}, Sweden}
\date{\today}

\begin{abstract}
To better understand the inner workings of information spreading, network researchers often use simple models to capture the spreading dynamics.
But most models only highlight the effect of local interactions on the global spreading of a single information wave, and ignore the effects of interactions between multiple waves.
Here we take into account the effect of multiple interacting waves by using an agent-based model in which the interaction between information waves is based on their novelty.
We analyzed the global effects of such interactions and found that information that actually reaches nodes reaches them faster. This effect is caused by selection between information waves: slow waves die out and only fast waves survive.
As a result, and in contrast to models with non-interacting information dynamics,
the access to information decays with the distance from the source.
Moreover, when we analyzed the model on various synthetic and real spatial road networks, we found that the decay rate also depends on the path redundancy and the effective dimension of the system.
In general, the decay of the information wave frequency as a function of distance from the source follows a power law distribution with an exponent between -0.2 for a two-dimensional system with high path redundancy and -0.5 for a tree-like system with no path redundancy. We found that the real spatial networks provide an infrastructure for information spreading that lies in between these two extremes. Finally, to better understand the mechanics behind the scaling results, we provide analytical calculations of the scaling for a one-dimensional system.
\end{abstract}

\maketitle

\section {Introduction}
In today's society, we are flooded with information. Waves of new ideas, innovations, products, and trends follow each other in quick succession.
To better understand the inner workings of the dynamics, researchers often use simple models to capture important spreading mechanisms \cite{Valente199669,bikhchandani1998learning,TheoryOfFads,Kempe,goldenberg2001using,hedetniemi1988survey,bornholdt2011emergence} on a complex network \cite{RevModPhys.74.47,newman2003structure,boccaletti06,Sales-Pardo25092007,ClausetEtAl2008a,VespignaniNPhys2011,Jeong2000,Kleinberg:2000p5066,Milo2002}. Broadly speaking, there are two classes of such models: threshold models \cite{granovetter1978threshold,watts2002simple,bailey1975mathematical,hethcote2000mathematics,karimi2012threshold} and contagion models \cite{daley1964epidemics,PhysRevLett.92.218701,dodds2005generalized,nekovee2007theory}. Threshold models assume that individuals adopt new information or technology if a certain proportion of their friends have adopted it. This mechanism leads to cascades that, under favorable conditions, can propagate throughout the entire system. Contagion models assume that individuals spread information or rumors much like they spread microbial infections, through interactions. This mechanism can also cause spreading across the entire system, provided that the transmission rate is sufficiently high. Both types of models highlight the effect of local interactions on global spreading, but, in general, they ignore effects of interactions between multiple information waves.

Ideas inherently depend on each other, and waves of new information or technology often interact with one another as they propagate through society. In some systems, information waves integrate or hybridize, while in other systems they compete and replace one another. Here we focus on the latter type of interaction, when waves replace one another entirely, and analyze the global effects of such interactions.
For simplicity, we use novelty as a proxy for quality and key trait in the competition between waves \cite{rosvall2003modeling,lizana2010time}. Relevant systems include news media reporting of a particular event, release of new software versions, and invention of new technology that makes old technologies obsolete. With interaction between multiple waves, some waves will make it across the system and others will not. Therefore, the wave frequency, or, equivalently, the rate of adoption of individuals, will depend on their position relative to the information source in the system. For example, in ancient times, new methods of metallurgy spread in multiple waves across Europe, and, in modern times, new versions of operating systems spread across the globe. Not everybody upgrades immediately, and our aim in this paper is to analyze how the access to new information depends on the position in a system and the topology of the system.

To analyze the effects of interactions between multiple waves, we use a simple agent-based model introduced in ref.\ \cite{lizana2011modeling} and further analyzed in ref.\ \cite{PhysRevE.85.056116}. In its simplest formulation, a source agent injects multiple waves of new information over time in a given network. At a given rate, neighboring agents adopt the information if it is newer than the information that they already have. We provide analytical results of the wave frequency for a one-dimensional model and use simulations on lattice models between one and two dimensions, as well as on real spatial networks. In this way, we can quantify the effects of interactions between multiple waves and show, for example, that not only the distance from the source, but also the path redundancy, determine the rate of adoption. Moreover, compared to a system with non-interacting waves, new information reaches agents faster, because of selection between waves: slow waves die out and fast waves survive.

We begin by describing the model in detail and then analyze the model in different spatial geometries. In turn, we analyze the model from the perspective of the agents and the information waves, respectively.

\section {Methods}

In this section, we first detail the model and then describe the spatial embedding we use to analyze the spreading dynamics.

\subsection{Model definition}\label{ModDef}

The model consists of a number of agents, each of which occupies a node connected to neighboring nodes in a spatial network.
The core of the model describes interactions between multiple information waves released at a single source node.
At each time $t$, the source node in the center $j=0$ generates a new piece of information tagged by the time when it was generated $a_0(t) = t$.
In the same time step, each node $j$ with information of age $a_j(t)$ asks each of its neighbors $k_j$ with probability $\beta$ if $k_j$ has newer information. If $a_{k_{j}}(t) < a_j(t)$, $j$ adopts the new information and sets $a_j(t) = a_{k_{j}}(t)$. Without loss of generality and throughout our analysis, we use $\beta=0.5$.
Note that this model formulation is equivalent to one in which agents actively transmit information to each of their neighbors with probability $\beta$, and the neighbors update their information if it is newer than the information they already had. Therefore, if there was only one information wave or if the waves did not interact, the model would describe the standard SI dynamics of susceptible and infected individuals \cite{anderson1992infectious,hethcote2000mathematics,newman2010Intro}, and an information wave would always spread across the system. In the presence of multiple interacting waves, however, the information waves will compete with each other as they spread through the system, and only the fast ones will survive and make it across the system.

\begin{figure}[!ht]
\begin{center}
\includegraphics[width=0.995\columnwidth]{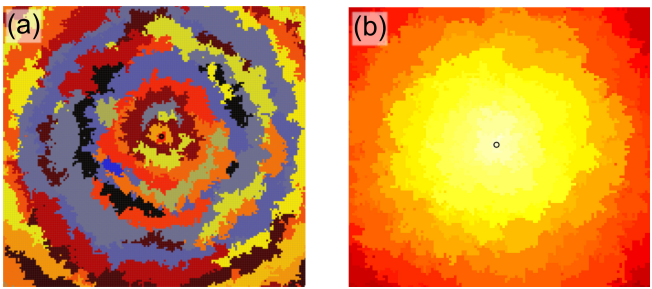}
\end{center}
\caption{{\bf The spatial and temporal dynamics of the spreading model with multiple interacting waves on a two-dimensional lattice.} \textbf{(a)} A snapshot of the dynamics in which each color corresponds to a single information wave. \textbf{(b)} The age landscape of the model  in which bright colors (light yellow) represent new information and dark colors (dark red) represent old information.
}
\label{Snap}
\end{figure}

Figure \ref{Snap} illustrates the dynamics of the spreading model on a two-dimensional lattice.
This figure was produced from the Java applet available in ref.\ \cite{Lizana2010Online}.
The source of new information is in the center of the lattice. Close to the information source, the diversity of information waves and the competition between them are high. Consequently, agents in this area become updated with high frequency. But far from the source, the wave frequency is lower, because high competition close to the source eliminates some waves.
Therefore, nodes in distant areas must wait longer between each update of information.
 For example, for a line source that is located at one edge of a lattice, the density of wave fronts decays as the square root of distance from the source \cite{lizana2011modeling}.
In this paper, we analytically derive this result for a one-dimensional system and further show that the information wave frequency also depends on the path redundancy, the number of shortest paths between the source and a given node.
The path redundancy can be thought of as the effective dimensionality of the system.
Higher path redundancy in a system gives nodes better access to new information.

\subsection{Spatial embedding}

\begin{figure}[!ht]
\begin{center}
\includegraphics[width=0.95\columnwidth]{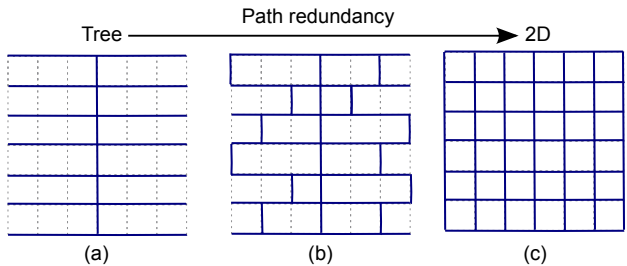}
\end{center}
\caption{{\bf } \textbf{Illustration of spatial networks with different degrees of path redundancy.} \textbf{(a)} A tree with path redundancy $\mathcal{R}=0$. \textbf{(b)} A network with path redundancy $\mathcal{R}=0.33$. \textbf{(c)} A two-dimensional lattice with path redundancy $\mathcal{R}=1$}
\label{schem}
\end{figure}

To analyze the effects of path redundancy, we build synthetic spatial networks with varying degree of path redundancy. The networks range from trees to two-dimensional lattices (Fig.~\ref{schem}).
In the two-dimensional lattice, the number of shortest paths grows exponentially as a function of distance from the source.
We construct the networks in two steps:
\begin{itemize}
\item[($i$)] We start with a two-dimensional structure with nodes connected in horizontal rows and one vertical column through the source node in the center (Fig.\ \ref{schem}(a)).
\item[($ii$)]We then randomly connect a fraction $\mathcal{R}$ of the remaining disconnected pairs of nodes (Fig.\ \ref{schem}(b,c)).
\end{itemize}
We quantify the path redundancy in terms of $\mathcal{R}$, where $\mathcal{R}=0$ corresponds to a tree and $\mathcal{R}=1$ corresponds to a two-dimensional lattice.
Figure \ref{schem} schematically shows how, by connecting disconnected pairs in the tree structure in Fig.\ \ref{schem}(a), we can increase the path redundancy through intermediate values in Fig.\ \ref{schem}(b) to high values in the fully connected two-dimensional lattice in Fig.\ \ref{schem}(c).

To complement the analysis of synthetic networks, we also analyze two real spatial networks with effective dimensionality between one and two: the road networks of Texas and California \cite{SNAP}. For all of the described networks, we quantify the wave frequency as a function of distance and path redundancy. For comparison, we compare the results with a null model without interactions between information waves. For the one- and two-dimensional systems, we also quantify the wave-speed distribution as a function of distance and path redundancy.

\begin{figure*}[!htb]
\begin{center}
\includegraphics[width=1.95\columnwidth]{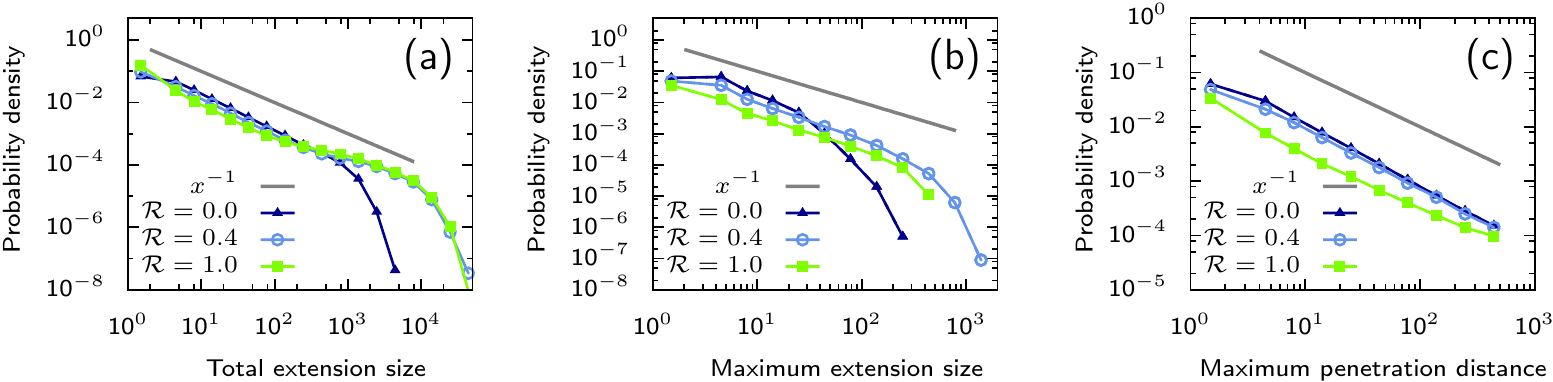}
\end{center}
\caption{{\bf Spatial spreading profile of competing information waves on different networks.} The networks have $1.2 \cdot 10^6$ nodes and each point corresponds to an average over more than 5 independent runs. Each run simulates competition between 15,000 different information waves. For the total extension size in (a), we normalize by the number of waves that die out within the system, and for the maximum extension size in (b) and the maximum penetration distance in (c), we normalize by the number of released waves.
}
\label{Cum_CS}
\end{figure*}

\section{Results and Discussion}

Before we show the results for the speed and wave frequency as a function of distance and path redundancy, we begin by analyzing the dynamics from the information waves' perspective.

\subsection{Spatial spreading profile}

Unlike non-interacting information waves, many interacting information waves will die out long before they reach the boundaries of the system.
For a general idea of how far they spread, we analyzed the spatial spreading profile by measuring the probability distribution of the information waves' total extension, maximum extension, and maximum penetration distance. The total extension of an information wave over its entire lifetime is the fraction of nodes in the network that, at some point, adopted the corresponding information. This measure captures the aggregated popularity of a piece of information over its entire lifetime. The maximum extension of an information wave at its peak is the maximum fraction of nodes in the network that simultaneously adopted the corresponding information as their latest information. This measure reflects the maximum popularity of a piece of information over its whole lifetime. Finally, the maximum penetration distance of a wave is the longest geodesic distance from the source that the wave ever reached. 

Figure \ref{Cum_CS} shows the spatial spreading profiles of interacting waves on three networks with different levels of path redundancy. We used a path redundancy of 0.4 as an intermediate value, because, as we show in the next section, a path redundancy of $\mathcal{R}=0.4$ corresponds to the average path redundancy of the road networks of Texas and California.
As Fig.\ \ref{Cum_CS} shows, the dynamic behavior of this intermediate path redundancy is similar to the maximum path redundancy of the two-dimensional lattice. For all topologies, the competition between waves is most intense close to the source, such that most waves die out small before they have covered much ground (Fig.\ \ref{Cum_CS}(a)). Except for boundary effects, which are significant in some cases, a power law distribution with exponent 1 approximately captures the scaling for all topologies. While the scaling is similar for different degrees of path redundancy, higher path redundancy increases the overall probability of spreading across the system. That is, in a topology with higher path redundancy and more possibilities for a wave to escape from chasing waves, more waves can reach the system boundary. As a result, the fraction of waves that die before reaching the boundary is smaller in a system with higher path redundancy, as seen by the vertically shifted probability densities between high and low path redundancy in Figs.\ \ref{Cum_CS}(b) and (c). With no path redundancy in a tree-like topology, there is only one direction to expand into and chasing waves follow immediately after. Therefore, very few waves occupy many nodes at the same time in a low-dimensional system (Fig.\ \ref{Cum_CS}(b)). In contrast, with higher path redundancy, a fast wave can expand more quickly in multiple directions and reach higher maximum extension size.

Interactions between information waves prevent slow waves from reaching distant parts of the network. For the individuals that propel the spread of the information, this competition affects ($i$) the age of the information that actually reaches them, and ($ii$) the frequency at which new information arrives. In the next section we take the perspective of the waves and, in turn, investigate these two effects in detail.

\subsection{Information wave speed and frequency}

\begin{figure}[!ht]
\begin{center}
\includegraphics[width=0.8\columnwidth]{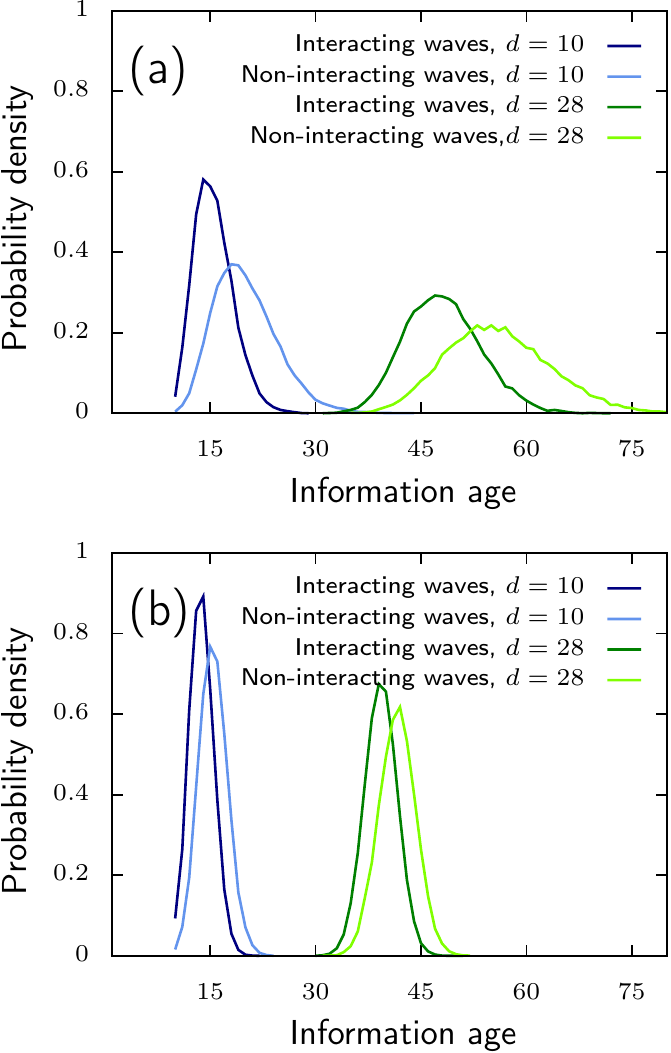}
\end{center}
\caption{\label{AgeFig}
{\bf Probability density of information age.}
\textbf{(a)} On a tree with 3600 nodes and $\mathcal{R}=0$.
\textbf{(b)} On a two-dimensional lattice with 3600 nodes and $\mathcal{R}=1$.
We quantified the age distribution of arriving waves for nodes that are close to the source ($d=10$) and for nodes that are farther away from the source ($d=28$). Results are obtained by averaging over more than 10,000 different competing waves.
}
\end{figure}

To investigate the extent of novelty for information arriving at a node, we calculated the age distribution of waves that reach a certain area.
That is, we measured the average age of hitting waves for nodes at a given geodesic distance $d$ to the source.
For comparison, we did the same experiment for multiple non-interacting waves.
We ran our experiment on a network with 3600 nodes (ordered in a 60 by 60 square) and quantified the probability distribution of information age for two groups of nodes: those that are close to the source ($d=10$) and those that are far from the source ($d=28$).
In Fig.~\ref{AgeFig}, we compare the probability distribution of the information age between interacting and non-interacting waves on two networks: a tree with the lowest possible path redundancy, $\mathcal{R}=0$, and a two-dimensional lattice with the highest possible path redundancy, $\mathcal{R}=1$.
In both networks, information packages that reach a node have traveled for a shorter time in systems with interacting waves, because the interaction between waves forms a selection process in which only fast waves survive.

\begin{figure}[!ht]
\begin{center}
\includegraphics[width=0.8\columnwidth]{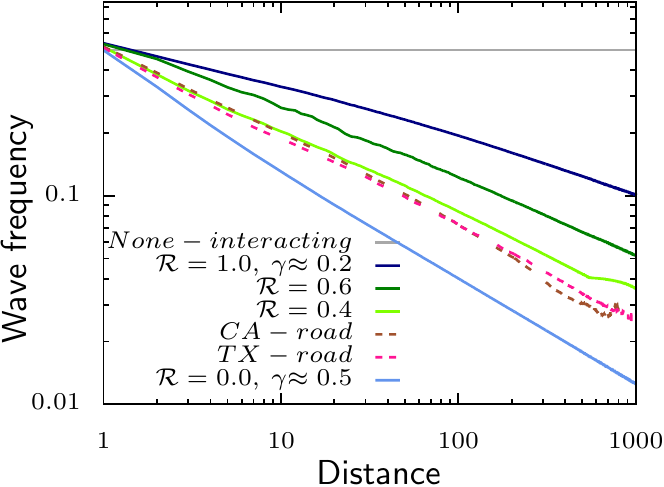}
\end{center}
\caption{
{\label{Road} \bf Wave frequency as a function of distance for interacting and non-interacting information waves on different networks.}
The results on the tree and two-dimensional lattice are fitted to a power-law with exponent 0.5 and 0.3, respectively.
For the California and Texas road networks, the results are very close to each other and similar to the network with path redundancy $\mathcal{R}=0.4$.
All synthetic networks have more than $10^6$ nodes. The results on these networks are achieved by simulating more than 5000 competing waves. The results on the road networks are averaged over more than 25 runs and each run includes more than 30,000 competing waves.}
\end{figure}

Information that reaches a node is newer with than without interaction between waves, because slow waves die out as they move away from the source. That is, nodes far from the source will only be reached by a fraction of all pieces of information that spread from the source. We quantified this effect with the wave frequency, the rate at which new information waves arrive at a node. In Fig.\ \ref{Road}, we show how the wave frequency scales as a function of the distance from the source. We quantified the wave frequency as a function of geodesic distance on multiple networks: a tree  ($\mathcal{R}=0$) with 1,210,000 nodes (1100 by 1100 square), a two-dimensional lattice ($\mathcal{R}=1.0$) with the same number of nodes,
and two synthetic networks with the same number of nodes and path redundancies between the tree and the lattice: one with $\mathcal{R}=0.4$ and one with $\mathcal{R}=0.6$.

For non-interacting waves, the wave frequency is the same for any node at any location and equal to the transition probability $\beta$ (gray line). For interacting waves, the wave frequency decays as a power law of the form $f(d) \sim d^{-\gamma}$ with an exponent $\gamma$ that depends on the path redundancy. In general, for nodes at similar geodesic distance, a topology with higher path redundancy results in higher wave frequency. For example, the wave frequency decays similarly fast for the two road networks and close to the synthetic network with path redundancy $\mathcal{R}=0.4$. Moreover, the exponent $\gamma$ is around 0.2 for the two-dimensional lattice and around 0.5 for the tree structure with no path redundancy.


Historically, the spread of technology and innovation has taken place on spatial networks, such as road and river networks. Trade routes, such as the Silk Road connecting the East and the West, worked as the backbone of the spread of information for centuries. To capture the dynamics on such networks, and on similar networks with intermediate path redundancies, we analyzed the California and Texas road networks. We found that the dynamics on the spatial networks are similar to the synthetic networks with path redundancy $\mathcal{R}=0.4$; the access to new information as a function of distance from the source has a power-law scaling (Fig.\ \ref{Road}). To better understand the origin of this universal power-law behavior, in the next section we derive an analytical calculation for the wave frequency in a one-dimensional system.

\subsection{Analytical derivation of the wave frequency}

\begin{figure}[!thb]
\begin{center}
\includegraphics[width=0.85\columnwidth]{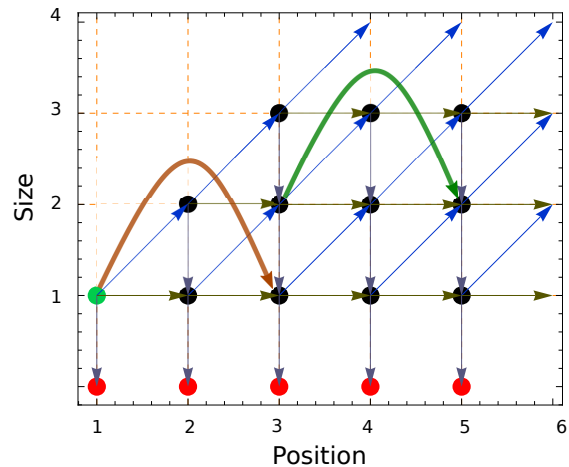} 
\end{center}
\caption{  {\bf  State transition diagram of position and size}
The position and size of the waves can change by following this diagram's arrows. The horizontal axis represents the position $r$ of the right boundary, and the vertical axis represents the size $s$ of the wave.}
\label{stateWS}
\end{figure}

In this section, we provide an analytical derivation of the wave-frequency scaling in a one-dimensional system. We derive the wave frequency as a function of distance from the source by working with two quantities: the wave size $s$ and the position $r$ of the outer boundary of the wave, or ``position'' for short. In one dimension,  these two magnitudes are really governed by what happens in the two boundaries in a single time step. Specifically, the following outcomes are possible: 

\begin{itemize}
    \item{
        The two boundaries move simultaneously to the right; this happens with probability $\beta^2$. In that case, $s$ remains the same and $r$ increases. This corresponds to the thin horizontal arrows in Fig.\ \ref{stateWS}.}
    \item{
        The inner boundary remains in the same position, and the outer one moves. This implies that both $r$ and $s$ increase by one, which we represent with thin diagonal arrows in Fig.\ \ref{stateWS}. This happens with probability $\beta (1 - \beta)$.
    }
    \item{
        The inner boundary moves, and the outer remains in the same position. The size decreases by one and $r$ keeps the previous value. This again happens with probability $\beta (1-\beta)$, and we represent it with thin downward arrows in Fig.\ \ref{stateWS}. 
    }
    \item {
        Both boundaries remain in the same position, in which case neither the wave's position nor its size change. This happens with probability $(1-\beta)^2$. 
    }
\end{itemize}

These probabilities sum up to one, but since we are just interested in size and position, and the fourth outcome changes none of these, we remove the fourth outcome and normalize the other probabilities accordingly. Thus the  probability for the first outcome becomes $\beta^2/\left( 1 - (1-\beta)^2 \right) = \beta/(2-\beta)$, which we denote by $b$. The second and third case  become $\beta (1-\beta) / \left( 1 - (1-\beta)^2 \right) = \beta (\beta-1) / (\beta - 2 )$. The second and third case have the same probability, but we denote them with different letters, $a$ and $c$ respectively, for the sake of clarity. Each of  the three kinds of transitions is depicted in figure \ref{stateWS} with arrows of different colors. 

In the model, the origin is special because a new wave starts there at each time step; no other points share that property. To make the analysis simpler, we will consider an alternative origin one step outwards, where the starting waves have size 1 with probability 1. The state corresponding to this new origin is represented with a green dot in the figure. Effectively, any information we obtain using this new origin is conditioned on the wave actually reaching this first point. We use this fact later to restore the original condition of a new wave at the origin of each time step. 

Using the transition probabilities $a$, $b$, and $c$, and guided by Fig.\ \ref{stateWS}, we can derive a recursive equation for the probability of having a size-one wave at a given position $r$. We use the fact that, in the state space represented in the diagram, the marginal probability of all the walks starting and ending with the same value for $s$, never going below $s$  and advancing $x$ steps, does not depend on $s$ itself, but only in the number of steps $x$ advanced. In other words, the thick curved arrows in the figure correspond to events with the same probability, even if they use different values for $s$. We call this probability $g(x)$. Note that then $g(r)$ is the probability of finding the wave at position $r$ with size 1. 

We can get an expression for $g(r)$ as follows. If a walk ends at  $(r,1)$, it can reach this last state in only  two ways: from the left, with a green arrow, or from the state above $(r,2)$, with a gray arrow. If the state is reached from the left, it means that the wave had to go from $(1,1)$ to $(r-1,1)$, and thus we need the value of $g(r-1)$. The case for the wave coming from above is more convoluted and we handle it in the following way. First, because the state above had to go to 2, we can safely assume that there was at least one transition where $s$ changed from 1 to 2. We fix the position of the last of such a transition as $r_1$, such that $(r_1,1)$ and $(r_1+1,2)$ is a segment of the walk. We can see that the walk from $(1,1)$ to $(r_1,1)$ will have probability $g(r_1)$.  The segment from $(r_1,1)$ to $(r_1+1,2)$ will have probability $a$. That leaves us with the part of the walk from $(r_1+1,2)$ to $(r,2)$. Because the last transition from $s=1$ to $s=2$ was the one at $r_1$, there is no way that the walk could go to $s<2$ in the section from $(r_1+1,2)$ to $(r,2)$. In other words,  this part has probability $g(r-r_1-1)$. With this information, we can write the recurrence for $g(r)$ as
\begin{equation} \label{eqM2}
g(r)=\sum _{r_1=0}^{r-1} \left[ g\left(r_1\right) a\text{  }g\left(r-r_1-1\right) c \right]+ b g(r-1).
\end{equation}
To solve the above equation, we use a generator function of the form
\begin{equation} \label{eqM3}
G(z)=\sum_{i=0}^{\infty}g(i)\, z^{i}.
\end{equation}
Since Eq.\ \ref{eqM2} is only valid for $r\geq2$, we first write
\begin{equation}\label{eqGz}
G(z)\text{=}g(0)+z\, g(1)+\sum_{i=2}^{\infty}g(i)\, z^{i},
\end{equation}
then apply the recursivity to obtain
\begin{equation} \label{eqM4}
\begin{split}
&G(z)=g(0)+z\, g(1)+\sum_{i=2}^{\infty}\, b\, g(i-1)\, z^{i}+
\\
&\sum_{i=2}^{\infty}z^{i}\sum_{i_{1}=0}^{-1+i}ac\, g(i-i_{1}-1)\, g(i_{1}).
\end{split}
\end{equation}
With some variable substitutions and algebraic manipulation, we can write it as
\begin{equation} \label{eqM5}
\begin{split}
&G(z)=g(0)-b\, z\, g(0)+z\, g(1)+b\, z\, G(z)+
\\
&ac\, z\sum_{j=0}^{\infty}z^{j}\sum_{i_{1}=0}^{j}g\left(j-i_{1}\right)g\left(i_{1}\right)-zac\, g(0)\, g(0).
\end{split}
\end{equation}
The terms in the sum of the previous expression represent a neat convolution, which can be expressed as the product of generating functions. From there we get the quadratic expression for $G(z)$,
\begin{equation}\label{eqM6}
G(z)=1-b\, z-ac\, z+(b+ac)\, z+b\, z\, G(z)+acz\, G(z)^{2}.
\end{equation}
From the two solutions, we select
\begin{equation}
G(z)=-\frac{-1+b\, z+\sqrt{1-2b\, z-4ac\, z+b^{2}z^{2}}}{2ac\, z}\label{eq:G-definition}.
\end{equation}

\begin{figure}[!tb]
\begin{center}
\includegraphics[width=0.95\columnwidth]{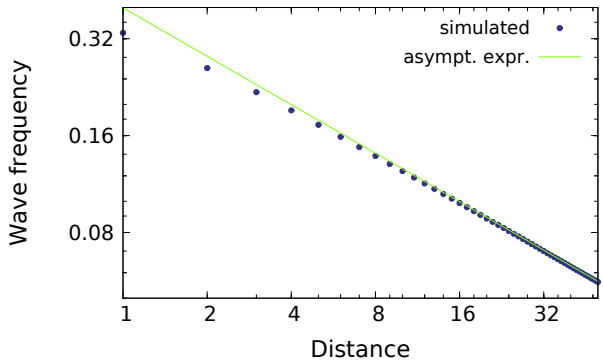} 
\end{center}
\caption{ {\bf The wave frequency at different positions $r$.} Dots: the frequency obtained averaging the observation of 2e6 time-steps. Green line: the theoretical prediction according to equation \ref{eq:frequency}. }
\label{SimIn1D}
\end{figure}

As mentioned in the text, with probability $c g(r)$, the wave is going to die without ever reaching the position $r + 1$. Thus, by summing $c g(x)$ from $x = 1$ to $x = r-1$, we can calculate the fraction of waves alive at a specific position $r$ of any size as
\begin{equation}
h(r)=1-c\sum\limits_{x=0}^{r-1} g(x).
\end{equation}
To calculate the survival probability $h(r)$, we write down the corresponding generating function as
\begin{equation}
H(z)=\sum_{i=0}^{\infty}z^{i}-\sum_{r_{0}=0}^{\infty}z^{i}c\sum_{r=0}^{i-1}g(r).
\end{equation}
The first term in the difference is $\frac{1}{1-z}$ and the second term is again a convolution:
\begin{equation}
H(z)\text{=}\frac{1}{1-z}-c\sum_{i=0}^{\infty}z^{i}\sum_{j=0}^{i}g(j)+c\sum_{i=0}^{\infty}z^{i}g(i),
\end{equation}
such that we get
\begin{equation}
H(z)=\frac{c\, z\, G(z)-1}{z-1}.
\end{equation}
After substituting Eq.\ \ref{eqGz} and doing some simplifications, we obtain
\begin{equation}
H(z)=\frac{2}{\sqrt{\left(-4+\beta\left(4+\beta\left(-1+z\right)\right)\right)\left(-1+z\right)}-\beta\,(z-1)}\label{eq:gen-for-H}.
\end{equation}
The function $H(z)$ has only one principal singularity at $z-1$, and we know that the coefficients $h(r)$ are strictly positive. Therefore, we can apply Corollary 2 in ref.\ \cite{flajolet1990singularity}
and derive the asymptotic scaling for $h(r)$. By that corollary, we get that, when $r\rightarrow\infty$,
\begin{equation}
h(r)\sim\frac{1}{\sqrt{1-\beta}\sqrt{\pi}}\, r^{-1/2}
\end{equation}
We should remember that $h(r)$ represents the survival probability
of the wave once it takes off at the first position inmediatly after the origin, and that happens with probability
$\beta$, so the frequency at which new waves are observed at a
given point $r$ is
\begin{equation}\label{eq:frequency}
f(r)\sim\frac{\beta}{\sqrt{1-\beta}\sqrt{\pi}}\, r^{-1/2}
\end{equation}
This expression is valid as long as $\beta<1$, provided that $r$ is sufficiently large. Figure \ref{SimIn1D} shows
the values of the frequency obtained by simulation and those obtained by the previous equation.
\bigskip

\section{Conclusion}
We used a simple agent-based model to capture the observation that waves of new information or technology often interact with one another as they propagate through a system. In the model, we use novelty as a proxy for quality and key trait in the interaction between waves. We showed that information that reaches agents is newer with than without interactions between waves at the cost of lower arrival frequency of information waves. Moreover, high path redundancy has a positive effect on the wave frequency, such that information more easily spreads in a system with multiple routes to targets. In general, the wave frequency decays as a power law of the distance from the source, and analytically we showed that the scaling goes as one over the square root of the distance in a one-dimensional system.
Our analysis on road networks of California and Texas showed that these networks provide an infrastructure for information propagation that corresponds to lattice models between one and two dimensions. We conclude that interacting information waves show interesting dynamics that call for further study.

\begin{acknowledgments}
We are grateful to Akhil Kedia for many valuable suggestions. This research was conducted using the resources of High Performance Computing Center North (HPC2N). Martin Rosvall was supported by the Swedish Research Council grant 2012-3729, and Ludvig Lizana wishes to acknowledge financial support from Knut and Alice Wallenberg foundation.
\end{acknowledgments}

\bibliography{TW_PRE_110913}

\begin{thebibliography}{34}%
\makeatletter
\providecommand \@ifxundefined [1]{%
 \@ifx{#1\undefined}
}%
\providecommand \@ifnum [1]{%
 \ifnum #1\expandafter \@firstoftwo
 \else \expandafter \@secondoftwo
 \fi
}%
\providecommand \@ifx [1]{%
 \ifx #1\expandafter \@firstoftwo
 \else \expandafter \@secondoftwo
 \fi
}%
\providecommand \natexlab [1]{#1}%
\providecommand \enquote  [1]{``#1''}%
\providecommand \bibnamefont  [1]{#1}%
\providecommand \bibfnamefont [1]{#1}%
\providecommand \citenamefont [1]{#1}%
\providecommand \href@noop [0]{\@secondoftwo}%
\providecommand \href [0]{\begingroup \@sanitize@url \@href}%
\providecommand \@href[1]{\@@startlink{#1}\@@href}%
\providecommand \@@href[1]{\endgroup#1\@@endlink}%
\providecommand \@sanitize@url [0]{\catcode `\\12\catcode `\$12\catcode
  `\&12\catcode `\#12\catcode `\^12\catcode `\_12\catcode `\%12\relax}%
\providecommand \@@startlink[1]{}%
\providecommand \@@endlink[0]{}%
\providecommand \url  [0]{\begingroup\@sanitize@url \@url }%
\providecommand \@url [1]{\endgroup\@href {#1}{\urlprefix }}%
\providecommand \urlprefix  [0]{URL }%
\providecommand \Eprint [0]{\href }%
\providecommand \doibase [0]{http://dx.doi.org/}%
\providecommand \selectlanguage [0]{\@gobble}%
\providecommand \bibinfo  [0]{\@secondoftwo}%
\providecommand \bibfield  [0]{\@secondoftwo}%
\providecommand \translation [1]{[#1]}%
\providecommand \BibitemOpen [0]{}%
\providecommand \bibitemStop [0]{}%
\providecommand \bibitemNoStop [0]{.\EOS\space}%
\providecommand \EOS [0]{\spacefactor3000\relax}%
\providecommand \BibitemShut  [1]{\csname bibitem#1\endcsname}%
\let\auto@bib@innerbib\@empty
\bibitem [{\citenamefont {Valente}(1996)}]{Valente199669}%
  \BibitemOpen
  \bibfield  {author} {\bibinfo {author} {\bibfnamefont {T.~W.}\ \bibnamefont
  {Valente}},\ }\href {\doibase 10.1016/0378-8733(95)00256-1} {\bibfield
  {journal} {\bibinfo  {journal} {Social Networks}\ }\textbf {\bibinfo {volume}
  {18}},\ \bibinfo {pages} {69 } (\bibinfo {year} {1996})}\BibitemShut
  {NoStop}%
\bibitem [{\citenamefont {Bikhchandani}\ \emph {et~al.}(1998)\citenamefont
  {Bikhchandani}, \citenamefont {Hirshleifer},\ and\ \citenamefont
  {Welch}}]{bikhchandani1998learning}%
  \BibitemOpen
  \bibfield  {author} {\bibinfo {author} {\bibfnamefont {S.}~\bibnamefont
  {Bikhchandani}}, \bibinfo {author} {\bibfnamefont {D.}~\bibnamefont
  {Hirshleifer}}, \ and\ \bibinfo {author} {\bibfnamefont {I.}~\bibnamefont
  {Welch}},\ }\href@noop {} {\bibfield  {journal} {\bibinfo  {journal} {The
  Journal of Economic Perspectives}\ }\textbf {\bibinfo {volume} {12}},\
  \bibinfo {pages} {151} (\bibinfo {year} {1998})}\BibitemShut {NoStop}%
\bibitem [{\citenamefont {Bikhchandani}\ \emph {et~al.}(1992)\citenamefont
  {Bikhchandani}, \citenamefont {Hirshleifer},\ and\ \citenamefont
  {Welch}}]{TheoryOfFads}%
  \BibitemOpen
  \bibfield  {author} {\bibinfo {author} {\bibfnamefont {S.}~\bibnamefont
  {Bikhchandani}}, \bibinfo {author} {\bibfnamefont {D.}~\bibnamefont
  {Hirshleifer}}, \ and\ \bibinfo {author} {\bibfnamefont {I.}~\bibnamefont
  {Welch}},\ }\href
  {http://ideas.repec.org/a/ucp/jpolec/v100y1992i5p992-1026.html} {\bibfield
  {journal} {\bibinfo  {journal} {Journal of Political Economy}\ }\textbf
  {\bibinfo {volume} {100}},\ \bibinfo {pages} {992} (\bibinfo {year}
  {1992})}\BibitemShut {NoStop}%
\bibitem [{\citenamefont {Kempe}\ \emph {et~al.}(2003)\citenamefont {Kempe},
  \citenamefont {Kleinberg},\ and\ \citenamefont {Tardos}}]{Kempe}%
  \BibitemOpen
  \bibfield  {author} {\bibinfo {author} {\bibfnamefont {D.}~\bibnamefont
  {Kempe}}, \bibinfo {author} {\bibfnamefont {J.}~\bibnamefont {Kleinberg}}, \
  and\ \bibinfo {author} {\bibfnamefont {E.}~\bibnamefont {Tardos}},\ }in\
  \href {\doibase 10.1145/956750.956769} {\emph {\bibinfo {booktitle}
  {Proceedings of the ninth ACM SIGKDD }}},\ \bibinfo {series and number} {KDD '03}\
  (\bibinfo  {publisher} {ACM},\ \bibinfo {address} {New York, NY, USA},\
  \bibinfo {year} {2003})\ pp.\ \bibinfo {pages} {137--146}\BibitemShut
  {NoStop}%
\bibitem [{\citenamefont {Goldenberg}\ \emph {et~al.}(2001)\citenamefont
  {Goldenberg}, \citenamefont {Libai},\ and\ \citenamefont
  {Muller}}]{goldenberg2001using}%
  \BibitemOpen
  \bibfield  {author} {\bibinfo {author} {\bibfnamefont {J.}~\bibnamefont
  {Goldenberg}}, \bibinfo {author} {\bibfnamefont {B.}~\bibnamefont {Libai}}, \
  and\ \bibinfo {author} {\bibfnamefont {E.}~\bibnamefont {Muller}},\
  }\href@noop {} {\bibfield  {journal} {\bibinfo  {journal} {Academy of
  Marketing Science Review}\ }\textbf {\bibinfo {volume} {9}},\ \bibinfo
  {pages} {1} (\bibinfo {year} {2001})}\BibitemShut {NoStop}%
\bibitem [{\citenamefont {Hedetniemi}\ \emph {et~al.}(1988)\citenamefont
  {Hedetniemi}, \citenamefont {Hedetniemi},\ and\ \citenamefont
  {Liestman}}]{hedetniemi1988survey}%
  \BibitemOpen
  \bibfield  {author} {\bibinfo {author} {\bibfnamefont {S.~M.}\ \bibnamefont
  {Hedetniemi}}, \bibinfo {author} {\bibfnamefont {S.~T.}\ \bibnamefont
  {Hedetniemi}}, \ and\ \bibinfo {author} {\bibfnamefont {A.~L.}\ \bibnamefont
  {Liestman}},\ }\href@noop {} {\bibfield  {journal} {\bibinfo  {journal}
  {Networks}\ }\textbf {\bibinfo {volume} {18}},\ \bibinfo {pages} {319}
  (\bibinfo {year} {1988})}\BibitemShut {NoStop}%
\bibitem [{\citenamefont {Bornholdt}\ \emph {et~al.}(2011)\citenamefont
  {Bornholdt}, \citenamefont {Jensen},\ and\ \citenamefont
  {Sneppen}}]{bornholdt2011emergence}%
  \BibitemOpen
  \bibfield  {author} {\bibinfo {author} {\bibfnamefont {S.}~\bibnamefont
  {Bornholdt}}, \bibinfo {author} {\bibfnamefont {M.~H.}\ \bibnamefont
  {Jensen}}, \ and\ \bibinfo {author} {\bibfnamefont {K.}~\bibnamefont
  {Sneppen}},\ }\href@noop {} {\bibfield  {journal} {\bibinfo  {journal} {Phys.
  Rev. Lett.}\ }\textbf {\bibinfo {volume} {106}},\ \bibinfo {pages} {058701}
  (\bibinfo {year} {2011})}\BibitemShut {NoStop}%
\bibitem [{\citenamefont {Albert}\ and\ \citenamefont
  {Barab\'asi}(2002)}]{RevModPhys.74.47}%
  \BibitemOpen
  \bibfield  {author} {\bibinfo {author} {\bibfnamefont {R.}~\bibnamefont
  {Albert}}\ and\ \bibinfo {author} {\bibfnamefont {A.-L.}\ \bibnamefont
  {Barab\'asi}},\ }\href@noop {} {\bibfield  {journal} {\bibinfo  {journal}
  {Rev. Mod. Phys.}\ }\textbf {\bibinfo {volume} {74}},\ \bibinfo {pages} {47}
  (\bibinfo {year} {2002})}\BibitemShut {NoStop}%
\bibitem [{\citenamefont {Newman}(2003)}]{newman2003structure}%
  \BibitemOpen
  \bibfield  {author} {\bibinfo {author} {\bibfnamefont {M.~E.~J.}\
  \bibnamefont {Newman}},\ }\href@noop {} {\bibfield  {journal} {\bibinfo
  {journal} {SIAM Rev.}\ }\textbf {\bibinfo {volume} {45}},\ \bibinfo {pages}
  {167} (\bibinfo {year} {2003})}\BibitemShut {NoStop}%
\bibitem [{\citenamefont {Boccaletti}\ \emph {et~al.}(2006)\citenamefont
  {Boccaletti}, \citenamefont {Latora}, \citenamefont {Moreno}, \citenamefont
  {Chavez},\ and\ \citenamefont {Hwang}}]{boccaletti06}%
  \BibitemOpen
  \bibfield  {author} {\bibinfo {author} {\bibfnamefont {S.}~\bibnamefont
  {Boccaletti}}, \bibinfo {author} {\bibfnamefont {V.}~\bibnamefont {Latora}},
  \bibinfo {author} {\bibfnamefont {Y.}~\bibnamefont {Moreno}}, \bibinfo
  {author} {\bibfnamefont {M.}~\bibnamefont {Chavez}}, \ and\ \bibinfo {author}
  {\bibfnamefont {D.-U.}\ \bibnamefont {Hwang}},\ }\href@noop {} {\bibfield
  {journal} {\bibinfo  {journal} {Phys. Rep.}\ }\textbf {\bibinfo {volume}
  {424}},\ \bibinfo {pages} {175} (\bibinfo {year} {2006})}\BibitemShut
  {NoStop}%
\bibitem [{\citenamefont {Sales-Pardo}\ \emph {et~al.}(2007)\citenamefont
  {Sales-Pardo}, \citenamefont {Guimerà}, \citenamefont {Moreira},\ and\
  \citenamefont {Amaral}}]{Sales-Pardo25092007}%
  \BibitemOpen
  \bibfield  {author} {\bibinfo {author} {\bibfnamefont {M.}~\bibnamefont
  {Sales-Pardo}}, \bibinfo {author} {\bibfnamefont {R.}~\bibnamefont
  {Guimerà}}, \bibinfo {author} {\bibfnamefont {A.~A.}\ \bibnamefont
  {Moreira}}, \ and\ \bibinfo {author} {\bibfnamefont {L.~A.~N.}\ \bibnamefont
  {Amaral}},\ }\href@noop {} {\bibfield  {journal} {\bibinfo  {journal} {Proc.
  Natl. Acad. Sci. USA}\ }\textbf {\bibinfo {volume} {104}},\ \bibinfo {pages}
  {15224} (\bibinfo {year} {2007})}\BibitemShut {NoStop}%
\bibitem [{\citenamefont {Clauset}\ \emph {et~al.}(2008)\citenamefont
  {Clauset}, \citenamefont {Moore},\ and\ \citenamefont
  {Newman}}]{ClausetEtAl2008a}%
  \BibitemOpen
  \bibfield  {author} {\bibinfo {author} {\bibfnamefont {A.}~\bibnamefont
  {Clauset}}, \bibinfo {author} {\bibfnamefont {C.}~\bibnamefont {Moore}}, \
  and\ \bibinfo {author} {\bibfnamefont {M.~E.~J.}\ \bibnamefont {Newman}},\
  }\href@noop {} {\bibfield  {journal} {\bibinfo  {journal} {Nature}\ }\textbf
  {\bibinfo {volume} {453}},\ \bibinfo {pages} {98} (\bibinfo {year}
  {2008})}\BibitemShut {NoStop}%
\bibitem [{\citenamefont {Vespignani}(2012)}]{VespignaniNPhys2011}%
  \BibitemOpen
  \bibfield  {author} {\bibinfo {author} {\bibfnamefont {A.}~\bibnamefont
  {Vespignani}},\ }\href@noop {} {\bibfield  {journal} {\bibinfo  {journal}
  {Nat Phys}\ }\textbf {\bibinfo {volume} {8}},\ \bibinfo {pages} {32}
  (\bibinfo {year} {2012})}\BibitemShut {NoStop}%
\bibitem [{\citenamefont {Jeong}\ \emph {et~al.}(2000)\citenamefont {Jeong},
  \citenamefont {Tombor}, \citenamefont {Albert}, \citenamefont {Oltvai},\ and\
  \citenamefont {Barab{\'a}si}}]{Jeong2000}%
  \BibitemOpen
  \bibfield  {author} {\bibinfo {author} {\bibfnamefont {H.}~\bibnamefont
  {Jeong}}, \bibinfo {author} {\bibfnamefont {B.}~\bibnamefont {Tombor}},
  \bibinfo {author} {\bibfnamefont {R.}~\bibnamefont {Albert}}, \bibinfo
  {author} {\bibfnamefont {Z.}~\bibnamefont {Oltvai}}, \ and\ \bibinfo {author}
  {\bibfnamefont {A.~L.}\ \bibnamefont {Barab{\'a}si}},\ }\href@noop {}
  {\bibfield  {journal} {\bibinfo  {journal} {Nature}\ }\textbf {\bibinfo
  {volume} {407}},\ \bibinfo {pages} {651} (\bibinfo {year}
  {2000})}\BibitemShut {NoStop}%
\bibitem [{\citenamefont {Kleinberg}(2000)}]{Kleinberg:2000p5066}%
  \BibitemOpen
  \bibfield  {author} {\bibinfo {author} {\bibfnamefont {J.}~\bibnamefont
  {Kleinberg}},\ }\href@noop {} {\bibfield  {journal} {\bibinfo  {journal}
  {Nature}\ }\textbf {\bibinfo {volume} {406}},\ \bibinfo {pages} {845}
  (\bibinfo {year} {2000})}\BibitemShut {NoStop}%
\bibitem [{\citenamefont {Milo}\ \emph {et~al.}(2002)\citenamefont {Milo},
  \citenamefont {Shen-Orr}, \citenamefont {Itzkovitz}, \citenamefont {Kashtan},
  \citenamefont {Chklovskii},\ and\ \citenamefont {Alon}}]{Milo2002}%
  \BibitemOpen
  \bibfield  {author} {\bibinfo {author} {\bibfnamefont {R.}~\bibnamefont
  {Milo}}, \bibinfo {author} {\bibfnamefont {S.}~\bibnamefont {Shen-Orr}},
  \bibinfo {author} {\bibfnamefont {S.}~\bibnamefont {Itzkovitz}}, \bibinfo
  {author} {\bibfnamefont {N.}~\bibnamefont {Kashtan}}, \bibinfo {author}
  {\bibfnamefont {D.}~\bibnamefont {Chklovskii}}, \ and\ \bibinfo {author}
  {\bibfnamefont {U.}~\bibnamefont {Alon}},\ }\href@noop {} {\bibfield
  {journal} {\bibinfo  {journal} {Science}\ }\textbf {\bibinfo {volume}
  {298}},\ \bibinfo {pages} {824} (\bibinfo {year} {2002})}\BibitemShut
  {NoStop}%
\bibitem [{\citenamefont {Granovetter}(1978)}]{granovetter1978threshold}%
  \BibitemOpen
  \bibfield  {author} {\bibinfo {author} {\bibfnamefont {M.}~\bibnamefont
  {Granovetter}},\ }\href@noop {} {\bibfield  {journal} {\bibinfo  {journal}
  {American Journal of Sociology}\ ,\ \bibinfo {pages} {1420}} (\bibinfo {year}
  {1978})}\BibitemShut {NoStop}%
\bibitem [{\citenamefont {Watts}(2002)}]{watts2002simple}%
  \BibitemOpen
  \bibfield  {author} {\bibinfo {author} {\bibfnamefont {D.~J.}\ \bibnamefont
  {Watts}},\ }\href@noop {} {\bibfield  {journal} {\bibinfo  {journal} {Proc.
  Natl. Acad. Sci. USA}\ }\textbf {\bibinfo {volume} {99}},\ \bibinfo {pages}
  {5766} (\bibinfo {year} {2002})}\BibitemShut {NoStop}%
\bibitem [{\citenamefont {Bailey}(1975)}]{bailey1975mathematical}%
  \BibitemOpen
  \bibfield  {author} {\bibinfo {author} {\bibfnamefont {N.~T.}\ \bibnamefont
  {Bailey}},\ }\href@noop {} {\emph {\bibinfo {title} {The mathematical theory
  of infectious diseases and its applications}}}\ (\bibinfo  {publisher}
  {Charles Griffin and Company Ltd, 5a Crendon Street, High Wycombe, Bucks HP13
  6LE.},\ \bibinfo {year} {1975})\BibitemShut {NoStop}%
\bibitem [{\citenamefont {Hethcote}(2000)}]{hethcote2000mathematics}%
  \BibitemOpen
  \bibfield  {author} {\bibinfo {author} {\bibfnamefont {H.~W.}\ \bibnamefont
  {Hethcote}},\ }\href@noop {} {\bibfield  {journal} {\bibinfo  {journal} {SIAM
  review}\ }\textbf {\bibinfo {volume} {42}},\ \bibinfo {pages} {599} (\bibinfo
  {year} {2000})}\BibitemShut {NoStop}%
\bibitem [{\citenamefont {Karimi}\ and\ \citenamefont
  {Holme}(2012)}]{karimi2012threshold}%
  \BibitemOpen
  \bibfield  {author} {\bibinfo {author} {\bibfnamefont {F.}~\bibnamefont
  {Karimi}}\ and\ \bibinfo {author} {\bibfnamefont {P.}~\bibnamefont {Holme}},\
  }\href@noop {} {\bibfield  {journal} {\bibinfo  {journal} {arXiv preprint
  arXiv:1207.1206}\ } (\bibinfo {year} {2012})}\BibitemShut {NoStop}%
\bibitem [{\citenamefont {Daley}\ and\ \citenamefont
  {Kendall}(1964)}]{daley1964epidemics}%
  \BibitemOpen
  \bibfield  {author} {\bibinfo {author} {\bibfnamefont {D.~J.}\ \bibnamefont
  {Daley}}\ and\ \bibinfo {author} {\bibfnamefont {D.~G.}\ \bibnamefont
  {Kendall}},\ }\href@noop {} {\bibfield  {journal} {\bibinfo  {journal}
  {Nature}\ } (\bibinfo {year} {1964})}\BibitemShut {NoStop}%
\bibitem [{\citenamefont {Dodds}\ and\ \citenamefont
  {Watts}(2004)}]{PhysRevLett.92.218701}%
  \BibitemOpen
  \bibfield  {author} {\bibinfo {author} {\bibfnamefont {P.~S.}\ \bibnamefont
  {Dodds}}\ and\ \bibinfo {author} {\bibfnamefont {D.~J.}\ \bibnamefont
  {Watts}},\ }\href {\doibase 10.1103/PhysRevLett.92.218701} {\bibfield
  {journal} {\bibinfo  {journal} {Phys. Rev. Lett.}\ }\textbf {\bibinfo
  {volume} {92}},\ \bibinfo {pages} {218701} (\bibinfo {year}
  {2004})}\BibitemShut {NoStop}%
\bibitem [{\citenamefont {Dodds}\ and\ \citenamefont
  {Watts}(2005)}]{dodds2005generalized}%
  \BibitemOpen
  \bibfield  {author} {\bibinfo {author} {\bibfnamefont {P.~S.}\ \bibnamefont
  {Dodds}}\ and\ \bibinfo {author} {\bibfnamefont {D.~J.}\ \bibnamefont
  {Watts}},\ }\href@noop {} {\bibfield  {journal} {\bibinfo  {journal} {Journal
  of Theoretical Biology}\ }\textbf {\bibinfo {volume} {232}},\ \bibinfo
  {pages} {587} (\bibinfo {year} {2005})}\BibitemShut {NoStop}%
\bibitem [{\citenamefont {Nekovee}\ \emph {et~al.}(2007)\citenamefont
  {Nekovee}, \citenamefont {Moreno}, \citenamefont {Bianconi},\ and\
  \citenamefont {Marsili}}]{nekovee2007theory}%
  \BibitemOpen
  \bibfield  {author} {\bibinfo {author} {\bibfnamefont {M.}~\bibnamefont
  {Nekovee}}, \bibinfo {author} {\bibfnamefont {Y.}~\bibnamefont {Moreno}},
  \bibinfo {author} {\bibfnamefont {G.}~\bibnamefont {Bianconi}}, \ and\
  \bibinfo {author} {\bibfnamefont {M.}~\bibnamefont {Marsili}},\ }\href@noop
  {} {\bibfield  {journal} {\bibinfo  {journal} {Physica A: Statistical
  Mechanics and its Applications}\ }\textbf {\bibinfo {volume} {374}},\
  \bibinfo {pages} {457} (\bibinfo {year} {2007})}\BibitemShut {NoStop}%
\bibitem [{\citenamefont {Rosvall}\ and\ \citenamefont
  {Sneppen}(2003)}]{rosvall2003modeling}%
  \BibitemOpen
  \bibfield  {author} {\bibinfo {author} {\bibfnamefont {M.}~\bibnamefont
  {Rosvall}}\ and\ \bibinfo {author} {\bibfnamefont {K.}~\bibnamefont
  {Sneppen}},\ }\href@noop {} {\bibfield  {journal} {\bibinfo  {journal} {Phys.
  Rev. Lett.}\ }\textbf {\bibinfo {volume} {91}},\ \bibinfo {pages} {178701}
  (\bibinfo {year} {2003})}\BibitemShut {NoStop}%
\bibitem [{\citenamefont {Lizana}\ \emph {et~al.}(2010)\citenamefont {Lizana},
  \citenamefont {Rosvall},\ and\ \citenamefont {Sneppen}}]{lizana2010time}%
  \BibitemOpen
  \bibfield  {author} {\bibinfo {author} {\bibfnamefont {L.}~\bibnamefont
  {Lizana}}, \bibinfo {author} {\bibfnamefont {M.}~\bibnamefont {Rosvall}}, \
  and\ \bibinfo {author} {\bibfnamefont {K.}~\bibnamefont {Sneppen}},\
  }\href@noop {} {\bibfield  {journal} {\bibinfo  {journal} {Phys. Rev. Lett.}\
  }\textbf {\bibinfo {volume} {104}},\ \bibinfo {pages} {040603} (\bibinfo
  {year} {2010})}\BibitemShut {NoStop}%
\bibitem [{\citenamefont {Lizana}\ \emph {et~al.}(2011)\citenamefont {Lizana},
  \citenamefont {Mitarai}, \citenamefont {Sneppen},\ and\ \citenamefont
  {Nakanishi}}]{lizana2011modeling}%
  \BibitemOpen
  \bibfield  {author} {\bibinfo {author} {\bibfnamefont {L.}~\bibnamefont
  {Lizana}}, \bibinfo {author} {\bibfnamefont {N.}~\bibnamefont {Mitarai}},
  \bibinfo {author} {\bibfnamefont {K.}~\bibnamefont {Sneppen}}, \ and\
  \bibinfo {author} {\bibfnamefont {H.}~\bibnamefont {Nakanishi}},\ }\href@noop
  {} {\bibfield  {journal} {\bibinfo  {journal} {Phys. Rev. E}\ }\textbf
  {\bibinfo {volume} {83}},\ \bibinfo {pages} {066116} (\bibinfo {year}
  {2011})}\BibitemShut {NoStop}%
\bibitem [{\citenamefont {Dybiec}\ \emph {et~al.}(2012)\citenamefont {Dybiec},
  \citenamefont {Mitarai},\ and\ \citenamefont {Sneppen}}]{PhysRevE.85.056116}%
  \BibitemOpen
  \bibfield  {author} {\bibinfo {author} {\bibfnamefont {B.}~\bibnamefont
  {Dybiec}}, \bibinfo {author} {\bibfnamefont {N.}~\bibnamefont {Mitarai}}, \
  and\ \bibinfo {author} {\bibfnamefont {K.}~\bibnamefont {Sneppen}},\ }\href
  {\doibase 10.1103/PhysRevE.85.056116} {\bibfield  {journal} {\bibinfo
  {journal} {Phys. Rev. E}\ }\textbf {\bibinfo {volume} {85}},\ \bibinfo
  {pages} {056116} (\bibinfo {year} {2012})}\BibitemShut {NoStop}%
\bibitem [{\citenamefont {Anderson}\ \emph {et~al.}(1992)\citenamefont
  {Anderson}, \citenamefont {May},\ and\ \citenamefont
  {Anderson}}]{anderson1992infectious}%
  \BibitemOpen
  \bibfield  {author} {\bibinfo {author} {\bibfnamefont {R.~M.}\ \bibnamefont
  {Anderson}}, \bibinfo {author} {\bibfnamefont {R.~M.}\ \bibnamefont {May}}, \
  and\ \bibinfo {author} {\bibfnamefont {B.}~\bibnamefont {Anderson}},\
  }\href@noop {} {\emph {\bibinfo {title} {Infectious diseases of humans:
  dynamics and control}}},\ Vol.~\bibinfo {volume} {28}\ (\bibinfo  {publisher}
  {Wiley Online Library},\ \bibinfo {year} {1992})\BibitemShut {NoStop}%
\bibitem [{\citenamefont {Newman}(2010)}]{newman2010Intro}%
  \BibitemOpen
  \bibfield  {author} {\bibinfo {author} {\bibfnamefont {M.~E.~J.}\
  \bibnamefont {Newman}},\ }\href@noop {} {\emph {\bibinfo {title} {Networks:
  An Introduction}}}\ (\bibinfo  {publisher} {OxfordUniversity Press},\
  \bibinfo {address} {Oxford},\ \bibinfo {year} {2010})\BibitemShut {NoStop}%
\bibitem [{Liz()}]{Lizana2010Online}%
  \BibitemOpen
  \href@noop {} {}\bibinfo {note} {Java applet available here:
  \url{http://cmol.nbi.dk/models/japan/japanApplet.html}}\BibitemShut {NoStop}%
\bibitem [{SNA()}]{SNAP}%
  \BibitemOpen
  \href@noop {} {}\bibinfo {note} {Data from Stanford Large Network Dataset
  Collection available here: \url{http://snap.stanford.edu/data/}}\BibitemShut
  {NoStop}%
\bibitem [{\citenamefont {Flajolet}\ and\ \citenamefont
  {Odlyzko}(1990)}]{flajolet1990singularity}%
  \BibitemOpen
  \bibfield  {author} {\bibinfo {author} {\bibfnamefont {P.}~\bibnamefont
  {Flajolet}}\ and\ \bibinfo {author} {\bibfnamefont {A.}~\bibnamefont
  {Odlyzko}},\ }\href@noop {} {\bibfield  {journal} {\bibinfo  {journal} {SIAM
  Journal on discrete mathematics}\ }\textbf {\bibinfo {volume} {3}},\ \bibinfo
  {pages} {216} (\bibinfo {year} {1990})}\BibitemShut {NoStop}%
\end{thebibliography}%

\end{document}